# Large discrepancy between observations and simulations: Implications for urban air quality in China

Xiaokang Chen[1], Min Wang[1], Zhe Jiang[1]*, Yuqiang Zhang[2], Li Zhou[3]*, Jane Liu[4,5], Hong Liao[6], Helen Worden[7], Tai-Long He[8,9], Dylan Jones[8], Dongyang Chen[3], Qinwen Tan[10], Yanan Shen[1]

[1]School of Earth and Space Sciences, University of Science and Technology of China, Hefei, Anhui, 230026, China.
[2]Nicholas School of the Environment, Duke University, Durham, North Carolina, 27708, USA.
[3]College of Architecture and Environment, Sichuan University, Chengdu, Sichuan, 610000, China.
[4]School of Geographical Sciences, Fujian Normal University, Fuzhou, Fujian, 350007, China.
[5]Department of Geography and Planning, University of Toronto, Toronto, ON, M5S 3G3, Canada.
[6]School of Environmental Science and Engineering, Nanjing University of Information Science and Technology, Nanjing, Jiangsu, 210044, China.
[7]Atmospheric Chemistry Observations and Modeling Laboratory, National Center for Atmospheric Research, Boulder, CO, 80301, USA.
[8]Department of Physics, University of Toronto, Toronto, ON, M5S 1A7, Canada.
[9]Now at Department of Atmospheric Sciences, University of Washington, Seattle, WA 98195, USA.
[10]Chengdu Academy of Environmental Sciences, Chengdu, Sichuan, 610041, China.

*Correspondence to: Zhe Jiang (zhejiang@ustc.edu.cn) and Li Zhou (lizhou@scu.edu.cn)

## Abstract

Chemical transport models (CTMs) have been widely used to provide instructions for the control of ozone ($O_3$) pollution. However, we find large discrepancies between observation- and model-based urban $O_3$ chemical regimes: volatile organic compound (VOC)-limited regimes over N. China and weak nitrogen oxides ($NO_x$)-limited regimes over S. China in observations, in contrast to simulations with widespread distributions of strong $NO_x$-limited regimes. The conflicting $O_3$ evolutions are caused by underestimated urban $NO_x$ concentrations and the possible overestimation of biogenic VOC emissions. Reductions in $NO_x$ emissions, in response to regulations, have thus led to an unintended deterioration of $O_3$ pollution over N. China provinces, for example, an increase in surface $O_3$ by approximately 7 ppb over the Sichuan Basin (SCB) in 2014-2020. The $NO_x$-induced urban $O_3$ changes resulted in an increase in premature mortality by approximately 3000 cases in 2015-2020.

## 1. Introduction

Emission regulations have led to a significant decline in $NO_x$ emissions in China [*Jiang et al.*, 2022; *Zheng et al.*, 2018], accompanied by dramatic increases in surface $O_3$ concentrations [*Ma et al.*, 2019; *Sun et al.*, 2019; *Y-H Wang et al.*, 2020]. There are currently many efforts focusing on the opposite changes in surface $O_3$ and its precursors, with explanations including nonlinear $O_3$-$NO_x$ response [*Y Liu and Wang*, 2020; *Z Liu et al.*, 2021], the contribution from soil $NO_x$ emissions [*X. Lu et al.*, 2021], as well as changes in meteorological and fine particle ($PM_{2.5}$) conditions [*Dang et al.*, 2021; *K Li et al.*, 2019a; *K Li et al.*, 2019b]. A recent study [*Chen et al.*, 2021] further predicts mitigation of $O_3$ pollution over the highly industrialized North China Plain (NCP) and Yangtze River Delta (YRD) because $NO_x$ controls have led to a shift of $O_3$ chemical regimes from a VOC-limited to a transitional regime, and thus, stricter controls of $NO_x$ emissions are expected to lead to a decrease in $O_3$ pollution.

In contrast to highly industrialized E. China coast provinces, model studies suggested strong $NO_x$-limited regimes over inland China. Sustainable controls of $NO_x$ emissions are suggested to lead to decreases in $O_3$ pollution in recent years and are expected to lead to further decreases in $O_3$ pollution in the future. For example, *K Li et al.* [2019a] suggested that $NO_x$ reductions led to surface $O_3$ decrease over inland China by approximately 1 ppb/y in 2013-2017; *Y Liu and Wang* [2020] suggested that $NO_x$ reductions led to a surface $O_3$ decrease over inland China by approximately 3 ppb in 2013-2017; *Chen et al.* [2021] suggested that a 20% reduction in $NO_x$ emissions in 2019 will lead to an approximately 2-4 ppb decrease in surface $O_3$ over inland China; *Z Liu et al.* [2021] suggested that a 20% reduction in $NO_x$ emissions in 2016 will cause an up to 8 ppb decrease in surface $O_3$ over inland China. However, the hypothesis of strong $NO_x$-limited regimes over inland China conflicts with the observed $O_3$ increase.

The difficulty in explaining the observed surface $O_3$ changes over inland China implies possible deficiencies in our understanding of $O_3$ evolution, which poses a potential barrier to making effective regulatory policies to control $O_3$ pollution. In this study, we provide a comparative analysis of the observation- and model-based responses of urban $O_3$ concentrations to changes in $NO_x$ emissions in China in 2014-2020, by integrating surface in situ observations of $NO_2$, VOC and $O_3$, the GEOS-Chem chemical transport model (with 0.5°×0.625° horizontal resolution), and a photochemical box model. The objective of this work is to explore the causes for the observed $O_3$ changes and to evaluate the capability of state-of-the-art CTM to capture air quality evolution. The impact of long-term $O_3$ exposure is further analyzed based on the exposure-response function to understand their health effects.

## 2. Methods

### 2.1 MEE surface $NO_2$ and $O_3$ measurements

We use surface in situ $NO_2$ and $O_3$ concentration data from the China Ministry of Ecology and Environment (MEE) monitoring network for the period of 2014-2020. These real-time monitoring stations can report hourly concentrations of criteria pollutants from over 360 cities in 2019. Concentrations were reported by the MEE in units of ug/m$^3$ under standard temperature (273 K) until 31 August 2018. This reference state was changed on 1 September 2018 to 298 K. We converted the $O_3$ and $NO_2$ concentrations to ppb, and rescaled the postAugust 2018 concentrations to the standard temperature (273 K) to maintain the consistency in the trend analysis.

**2.2 GEOS-Chem model simulations**

The GEOS-Chem chemical transport model (http://www.geos-chem.org, version 12-8-1) is driven by assimilated meteorological data of MERRA-2 with nested 0.5°×0.625° horizontal resolution. The GEOS-Chem model includes fully coupled $O_3$-$NO_x$-VOC-halogen-aerosol chemistry. The chemical boundary conditions are updated every 3 hours from a global simulation with 4° × 5° resolution. Emissions in GEOS-Chem are computed by the Harvard-NASA Emission Component (HEMCO). We refer the reader to *Chen et al.* [2021] for the details of the model configurations.

**2.3 Photochemical box model**

The photochemical box model (OBM) is configured with master chemical mechanisms (MCM v3.3.11; http://mcm.york.ac.uk/home.htt). The MCM-OBM model was designed to investigate the atmospheric oxidation processes of VOC species [*X Liu et al.*, 2019; *Xue et al.*, 2013; *Xue et al.*, 2016]. The concentrations of sulfur dioxides ($SO_2$), carbon monoxide (CO), $NO_x$ and VOC, as well as meteorological parameters (atmospheric pressure, temperature, and relative humidity) from two monitoring sites in Chengdu city (in the SCB), were used as constraining parameters in the model. The MCM-OBM model simulations start at 12:00 local time for 8 hours, by inputting the observed $O_3$ concentration at the initial time. MCM-OBM simulations have been widely used to calculate the relative incremental reactivity to describe the response of $O_3$ to individual precursors [*He et al.*, 2019; *J Li et al.*, 2018; *Tan et al.*, 2018; *M Wang et al.*, 2020].

**2.4 Long-term health impacts of surface $O_3$ change**

We applied the following function to assess the mortality attributed to long-term $O_3$ exposure: $\triangle Health\ Impacts = y_0 \times (1 - e^{-\beta \triangle Exposure}) \times Population$, where $\triangle Health\ Impacts$ is the mortality change caused by the change in $O_3$ concentration in 2015-2020. $y_0$ is the baseline rates of specific diseases for various age groups, which we downloaded from the recent global burden of disease for 2019. $\beta$ is the exposure-response function for the unit of exposure. The relative risk (RR) value was set to 1.08 (90% confidence interval: 1.06, 1.11) based on previous studies [*Xiao Lu et al.*, 2020; *Turner et al.*, 2016; *Zhang et al.*, 2021]. In addition, we used the change in summertime MDA8 $O_3$ concentration in 2015-2020 (Fig. 4A) to represent $\triangle Exposure$. Population is the number of people in different age groups living in the same grid box (0.1º × 0.1º) obtained from the Gridded Population of the World database [2018] for 2019, and then re-gridded to 1°×1° resolution.

## 3. Results and Discussions

### 3.1 Observation-based $O_3$-$NO_2$ relationships

Fig. 1 shows the summertime maximum daily 8-hour average (MDA8) $O_3$ from the China Ministry of Ecology and Environment (MEE) monitoring network in 2019. Only urban stations are considered in this work, and background stations were excluded. The station-based measurements are averaged and regrided to 1°×1° resolution. The observed surface $O_3$ exhibits latitude dependence, i.e., higher $O_3$ in N. China and lower $O_3$ in S. China. Fig. 2A-B show the summertime $O_3$-$NO_2$ relationship over four domains in China at the MEE stations in the period of 2014-2020. The data (dots) are regional averages of MDA8 $O_3$ and $NO_2$ concentrations, binned into 1 ppb $NO_2$ increments. The lognormal fitting lines in Fig. 2A demonstrate nonlinear $O_3$-$NO_2$ relationships with turning points between $NO_x$- and VOC-limited regimes of approximately 9 ppb over the NCP, 10 ppb over the YRD, and 11 ppb over the PRD. We find shifts in $O_3$ chemical regimes from VOC-limited to transitional regimes over the NCP and YRD in 2014-2020. Our analysis suggests a $NO_x$-limited regime over the PRD in 2020 (Fig. 2A) and a VOC-limited regime over the SCB (Fig. 2B).

An observation-based photochemical box model (OBM) was further employed to evaluate the $O_3$ regime in the SCB. As shown in Fig. 2C, the OBM model indicates increases in $O_3$ with 60-80% reductions in $NO_x$ concentrations in Chengdu city (in the SCB) in 2017 and 2018. Similarly, Fig. 2B suggests an increase in $O_3$ with 40% reductions in $NO_2$ concentrations in the urban areas in the SCB in 2017 and 2018. The consistent VOC-limited regimes in the two observation-based approaches (Fig. 2B and Fig. 2C) support the feasibility of our approach (Fig. 2A-B) to analyze the $O_3$ chemical regime. The discrepancy between the two observation-based approaches (i.e., 60-80% and 40%) may reflect the difference in $O_3$ regimes among different cities in the SCB. As shown in Fig. 2B, reductions in $NO_x$ emissions have led to an increase in surface $O_3$ in the SCB by approximately 7 ppb in 2014-2020 (based on the lognormal fitting line). Stricter controls of $NO_x$ emissions and regardless of VOC are therefore expected to lead to even more $O_3$ pollution in the SCB.

## 3.2 Unintended deterioration of $O_3$ pollution over inland China

As shown in Fig. 2B, the lognormal fitting approach allows us to approximately predict the response of $O_3$ to $NO_2$ change. For example, the difference in the $O_3$ concentrations along the lognormal fitting line in Fig. 2B represents the predicted change in $O_3$ concentration ($dO_3$) in response to the increase in $NO_2$ concentrations ($dNO_2$) over the SCB. Furthermore, Fig. 3A shows the summertime $NO_2$ concentrations from the MEE stations in 2019. The 8-hour range of surface $NO_2$ measurements is selected according to the time range of MDA8 $O_3$, and then averaged and regrided to 1°×1° resolution. Following the prediction approach shown in Fig. 2B, lognormal fitting lines are produced for each 1°×1° grid in Fig. 3A, which allows us to predict grid-based $O_3$ changes in response to $NO_2$ in 2019.

Fig. 3B exhibits the response of surface $O_3$ to 20% increases in $NO_2$ concentrations in 2019. According to *Jiang et al.* [2022], we assume a 20% decline in anthropogenic $NO_x$ emissions in 2015-2019, and thus, Fig. 3B represents the observation-based $O_3$ changes driven by $NO_x$ changes in China in 2015-2019, approximately. We find widespread decreases in surface $O_3$ over N. China inland provinces in 2019 in response to 20% increases in $NO_2$ concentrations. In contrast to the hypothesis of strong $NO_x$-limited regimes [*Chen et al.*, 2021; *K Li et al.*, 2019a; *Y Liu and Wang*, 2020; *Z Liu et al.*, 2021], the observation-based analysis (Fig. 3B) indicates that $NO_x$ controls have led to increases in surface $O_3$ over N. China inland provinces. The 2018–2020 Chinese Clean Air Action plan called for a 9% decrease in $NO_x$ emissions [*CSC*, 2018], and thus, Fig. 3C further predicts surface $O_3$ changes driven by a 10% decrease in $NO_2$ concentrations. Continuous $NO_x$ controls, as shown in Fig. 3C, are thus predicted to result in the deterioration of $O_3$ pollution over N. China inland provinces.

**3.3 Discrepancy between observation- and model-based analyses**

Here we further investigate the $O_3$-$NO_2$ responses using the GEOS-Chem chemical transport model at 0.5°×0.625° horizontal resolution. Fig. 3D shows the summertime $NO_2$ concentrations from GEOS-Chem in 2019; Fig. 3E-F exhibit the modeled responses of surface $O_3$ to perturbations in anthropogenic $NO_x$ emissions (Run 1, See Table1). The modeled $NO_2$ and $O_3$ are sampled at the locations and times of the MEE surface measurements, and then averaged and regrided to 1°×1° resolution. Because only urban stations are considered in this work, the sampled simulations represent modeled urban $NO_2$ concentrations (Fig. 3D) and $O_3$ responses (Fig. 3E-F). The MEE

NO$_2$ observations are provided by chemiluminescence analyzers, in which NO$_2$ is catalytically transformed into nitrogen oxide (NO) by a molybdenum converter. Following *F Liu et al.* [2018], the modeled NO$_2$ concentrations in Fig. 3D are adjusted using the ratios of $NO_y/NO_2$. Here $NO_y = NO_2 + \Sigma AN + 0.95 \times PAN + 0.35 \times HNO_3$, where $\Sigma AN$ is the sum of all alkyl nitrate concentrations.

We find large discrepancies between the observations and simulations: 1) the sampled NO$_2$ concentrations (Fig. 3D) are comparable with surface NO$_2$ observations (Fig. 3A) over the industrialized NCP and YRD but lower in the rest of China. *F Liu et al.* [2018] found that modeled NO$_2$ concentrations are higher than surface NO$_2$ observations over industrialized NCP and YRD, and broadly lower in the rest of China. Different station types, i.e., all stations in *F Liu et al.* [2018] and urban stations in this work, as well as different periods, i.e., daily averages in *F Liu et al.* [2018] and MDA8 (based on O$_3$) in this work may affect the comparison between observations and simulations. For example, Fig. S1 shows large differences in the ratios of $NO_2/NO_y$ between different periods. 2) conflicting responses of O$_3$ to NO$_2$ changes between observation-based and model-based analyses in the rest of China: VOC-limited regimes (blue in Fig. 3B) over N. China and weak NO$_x$-limited regimes (slight red in Fig. 3B) over S. China in observations, in contrast to widespread distributions of strong NO$_x$-limited regimes in simulations (red in Fig. 3E).

**3.4 Impacts of biased urban NO$_2$ and VOC concentrations**

The underestimated urban NO$_2$ concentrations in Fig. 3D pose a significant barrier to simulating urban O$_3$ evolution. GEOS-Chem simulations (Run 2) adjust anthropogenic and soil NO$_x$ emissions over urban grids (i.e., grids have MEE urban stations) using the ratios of MEE/modeled NO$_2$. Emissions over the highly industrialized NCP, YRD and PRD are not adjusted because the modeled O$_3$ evolutions match well with observations in these areas [*Chen et al.*, 2021].

The adjustment of NO$_x$ emissions over urban grids led to enhancements of sampled NO$_2$ concentrations in Fig. 3G; however, they are still noticeably lower than observations (Fig. 3A), indicating influences from strong regional transport. Consequently, we further adjust regional background NO$_x$ emissions based on the ratios of averaged MEE/modeled NO$_2$ within neighboring grids (Run 3). As shown in Fig. 3J, the sampled NO$_2$ concentrations in Run 3 match well with the observed urban NO$_2$ concentrations (Fig. 3A). It should be noted that the adjustments of NO$_x$ emissions are designed to cover the influences of coarse model resolutions and strong regional transport on urban air quality simulations, which cannot be explained as underestimation in NO$_x$ emissions.

As shown in Fig. 3, the consistent NO$_2$ concentrations between the observations and simulations lead to significant improvements in the modeled urban O$_3$ evolution. For example, the observation-based analysis predicts a -1.8 ppb decrease in surface O$_3$ due to a 20% increase in anthropogenic NO$_x$ emissions in the SCB in 2019 (Fig. 3B); in contrast, the modeled responses are increases in surface O$_3$ by 2.8 ppb (Fig. 3E), 1.4 ppb (Fig. 3H) and 0.0 ppb (Fig. 3K). In a recent study, *Chen et al.* [2021] found that a 50% decrease in biogenic VOC emissions can improve surface O$_3$ simulations over the United States because of the reported overestimation of biogenic VOC emissions (MEGAN 2.1) in GEOS-Chem simulations [*Kaiser et al.*, 2018; *Wang et al.*, 2017]. Following *Chen et al.* [2021], GEOS-Chem simulations (Run 4) further reduce biogenic VOC emissions in China by 50%. The reductions in biogenic VOC emissions further improved the agreement between observations and simulations: the responses of surface O$_3$ to a 20% increase in anthropogenic NO$_x$ emissions are: -1.8 ppb (Fig. 3B) and -1.4 ppb (Fig. 3N) in the SCB.

Finally, Fig. 4A shows the predicted responses of O$_3$ concentrations (*d*O$_3$) to observed NO$_2$ changes (*d*NO$_2$) in 2015-2020. Changes in NO$_2$ concentrations have led to an increase of surface $O_3$ over provinces such as Anhui, Shanxi and Hebei (Fig. 4B). We then calculated the health impacts of long-term $O_3$ exposure. As shown in Fig. 4C, we find a noticeable increase in urban premature mortality burdens, by approximately 1500 cases in Anhui, and 500 cases in Zhe Jiang and Hebei Provinces. In contrast, changes in $O_3$ pollution led to a decrease in urban premature mortality burdens by approximately 1500 cases in Guangdong Province. It should be noted that our analysis is designed to predict urban $O_3$ pollution and health impacts (i.e., grids with urban MEE stations). The premature mortality burdens in this work are thus expected to be lower than those in the literature based on national total populations.

## 4. Conclusion

Chemical transport models, as powerful tools, have been widely used to provide instructions for the control of worldwide $O_3$ pollution. However, the comparative analysis in this work demonstrates large discrepancies between observation- and model-based urban $O_3$ chemical regimes: VOC-limited regimes over N. China; weak $NO_x$-limited regimes over S. China in the observations, in contrast to widespread distributions of strong $NO_x$-limited regimes in the simulations. The conflicting $O_3$ evolutions between observations and simulations are caused by underestimated urban $NO_x$ concentrations associated with coarse model resolutions and strong regional transport, as well as possible overestimation of biogenic VOC emissions. Different from the hypothesized decreases in surface $O_3$ driven by $NO_x$ control [*Chen et al.*, 2021; *K Li et al.*, 2019a; *Y Liu and Wang*, 2020; *Z Liu et al.*, 2021], reductions in $NO_x$ emissions have led to an unintended deterioration of $O_3$ pollution in 2014-2020 over N. China inland provinces. The $NO_2$-induced $O_3$ pollution changes resulted in an increase in premature mortality by approximately 3000 cases in 2015-2020. While $NO_x$ control is an effective pathway to mitigate $O_3$ pollution over S. China and E. coast provinces, our analysis highlights the importance of VOC controls for $O_3$ pollution mitigation over N. China inland provinces.

**Acknowledgments:**

The numerical calculations in this paper have been done on the supercomputing system in the Supercomputing Center of University of Science and Technology of China. This work was supported by the Hundred Talents Program of Chinese Academy of Science and National Natural Science Foundation of China (41721002). **Data Availability Statement:** We thank the China Ministry of Ecology and Environment for providing the surface $O_3$ and $NO_2$ measurements (from https://quotsoft.net/air/).

**Tables and Figures**

**Table. 1.** GEOS-Chem standard (Run 1) and sensitivity (Runs 2-4) simulations with 0.5°×0.625° horizontal resolution in Jun-Aug 2019. The scaling factors (A1) to enhance anthropogenic and soil $NO_x$ emissions over urban grids (i.e., grids have MEE urban stations) and urban grids+ regional background, based on the ratios between MEE and sampled $NO_2$ concentrations, are shown in Fig. S2. $NO_x$ (A1) and VOC (A3) emissions over highly industrialized NCP, YRD and PRD are not adjusted.

**Fig. 1.** Summertime MDA8 $O_3$ (with unit ppb) in 2019 from the MEE stations. The station-based measurements are averaged and re-grided to 1°×1° resolution. Only urban stations are considered. The black boxes define the domains used in this work. The star represents the location of Chengdu City in SCB.

**Fig. 2.** (A-B) Observed summertime $O_3$-$NO_2$ relationships from MEE stations with both $O_3$ and $NO_2$ measurements. The dots represent regional averages of MDA8 $O_3$ and $NO_2$ concentrations, binned into 1 ppb $NO_2$ increments. The 8-hour range of surface $NO_2$ measurements is selected according to the time range of MDA8 $O_3$. The lines are lognormal fitting lines. The error bars

represent standard error. The numbers (0 - 9) represent the summertime mean $O_3$ and $NO_2$ abundances, and a number itself corresponds a year with the year's last digit during 2014-2020. (C) 8-hour averaged responses of $O_3$ to $NO_x$ changes at the SL (103.93ºN, 30.58ºE, 20170801-20170815) and JPJ (104.05ºN, 30.66ºE, 20180601-20180610) sites in Chengdu, Sichuan Province. The simulations are performed using a photochemical box model with inputs of $SO_2$, CO, $NO_x$ and VOC concentrations and meteorological parameters from observations. Positive responses represent increases of $O_3$ within 8 hours (12-19 local time) due to a decrease of $NO_x$, indicating a VOC-limited regime.

**Fig. 3.** (A) Observed surface $NO_2$ concentrations from the MEE stations in Jun-Aug 2019. The station-based measurements are averaged and re-grided to 1°×1° resolution. (B-C) Predicted responses of MDA8 $O_3$ in 2019 to $NO_2$ changes based on the lognormal fitting lines. (D) Modeled surface $NO_2$ concentrations (Run1). The modeled $NO_2$ in panel D are adjusted using the ratios of $NO_y/NO_2$ to consider the influences from reactive oxidized nitrogen compounds in the chemiluminescence analyzers. (E-F) Modeled responses of MDA8 $O_3$ in 2019 to $NO_x$ emission changes. (G-O) Similar to panels D-F, but for sensitivity simulations (Runs 2-4). The modeled $NO_2$ and $O_3$ are sampled at the locations and times of MEE surface measurements, and then averaged and re-grided to 1°×1° resolution. The 8-hour range of surface $NO_2$ measurements is selected according to the time range of MDA8 $O_3$. The unit is ppb.

**Fig. 4.** (A-B) $NO_2$-induced urban $O_3$ changes in 2015-2020 based on the lognormal fitting method with unit ppb. (C) Changes in urban premature mortality burdens due to predicted $O_3$ changes.

| | | GEOS-Chem Simulations | | |
|---|---|---|---|---|
| | | Enhanced anthro + soil NOx (A1) | anthro NOx Factors (A2) | biogenic VOC Factors (A3) |
| Run1 (Standard) | #1 (Base) | N/A | 1.0 | 1.0 |
| | #2 | | 1.2 | 1.0 |
| | #3 | | 0.9 | 1.0 |
| Run2 | #1 (Base) | urban grids | 1.0 | 1.0 |
| | #2 | | 1.2 | 1.0 |
| | #3 | | 0.9 | 1.0 |
| Run3 | #1 (Base) | urban grids + regional backgrounds | 1.0 | 1.0 |
| | #2 | | 1.2 | 1.0 |
| | #3 | | 0.9 | 1.0 |
| Run4 | #1 (Base) | urban grids + regional backgrounds | 1.0 | 0.5 |
| | #2 | | 1.2 | 0.5 |
| | #3 | | 0.9 | 0.5 |

**Table. 1.** GEOS-Chem standard (Run 1) and sensitivity (Runs 2-4) simulations with 0.5°×0.625° horizontal resolution in Jun-Aug 2019. The scaling factors (A1) to enhance anthropogenic and soil NOx emissions over urban grids (i.e., grids have MEE urban stations) and urban grids+ regional background, based on the ratios between MEE and sampled NO₂ concentrations, are shown in Fig. S2. NOx (A1) and VOC (A3) emissions over highly industrialized NCP, YRD and PRD are not adjusted.

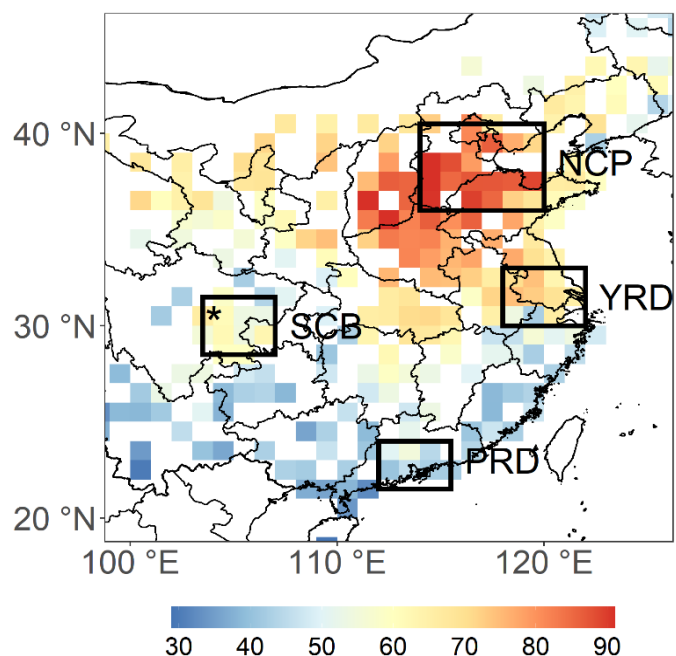

**Fig. 1.** Summertime MDA8 O₃ (with unit ppb) in 2019 from the MEE stations. The station-based measurements are averaged and re-gridded to 1°×1° resolution. Only urban stations are considered. The black boxes define the domains used in this work. The star represents the location of Chengdu City in SCB.

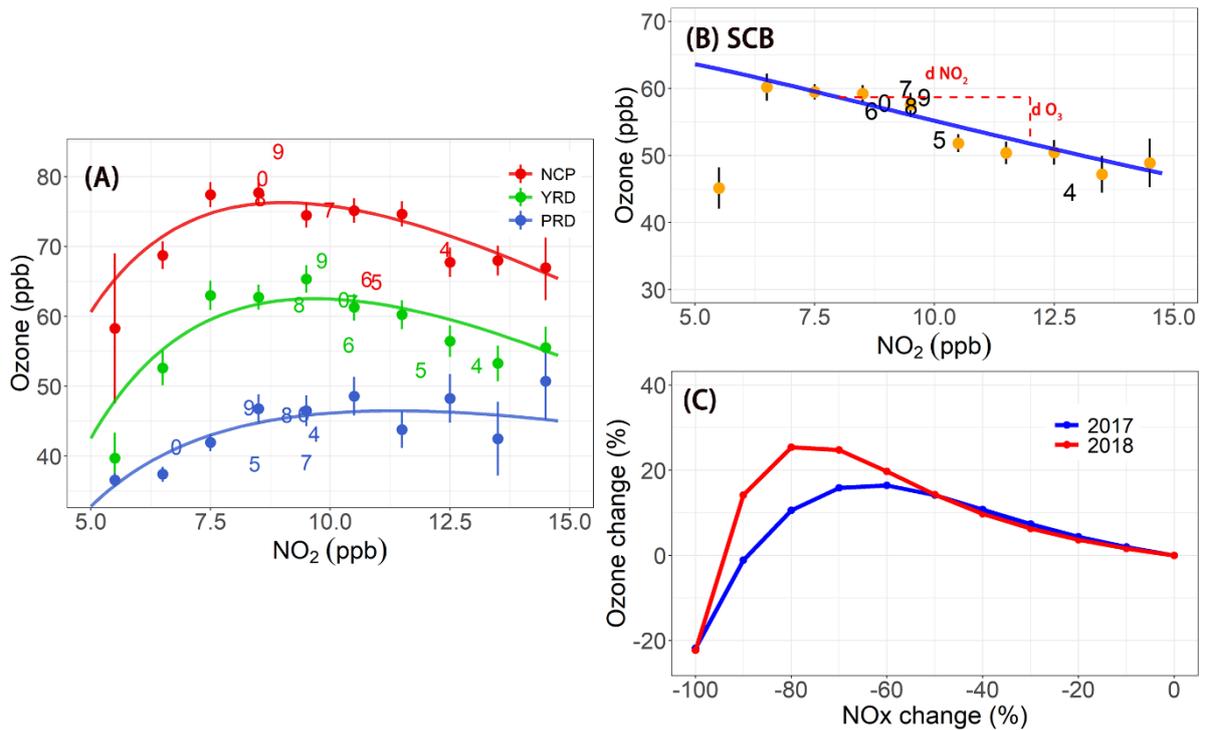

**Fig. 2.** (A-B) Observed summertime O$_3$-NO$_2$ relationships from MEE stations with both O$_3$ and NO$_2$ measurements. The dots represent regional averages of MDA8 O$_3$ and NO$_2$ concentrations, binned into 1 ppb NO$_2$ increments. The 8-hour range of surface NO$_2$ measurements is selected according to the time range of MDA8 O$_3$. The lines are lognormal fitting lines. The error bars represent standard error. The numbers (0 - 9) represent the summertime mean O$_3$ and NO$_2$ abundances, and a number itself corresponds a year with the year's last digit during 2014-2020. (C) 8-hour averaged responses of O$_3$ to NO$_x$ changes at the SL (103.93ºN, 30.58ºE, 20170801-20170815) and JPJ (104.05ºN, 30.66ºE, 20180601-20180610) sites in Chengdu, Sichuan Province. The simulations are performed using a photochemical box model with inputs of SO$_2$, CO, NO$_x$ and VOC concentrations and meteorological parameters from observations. Positive responses represent increases of O$_3$ within 8 hours (12-19 local time) due to a decrease of NO$_x$, indicating a VOC-limited regime.

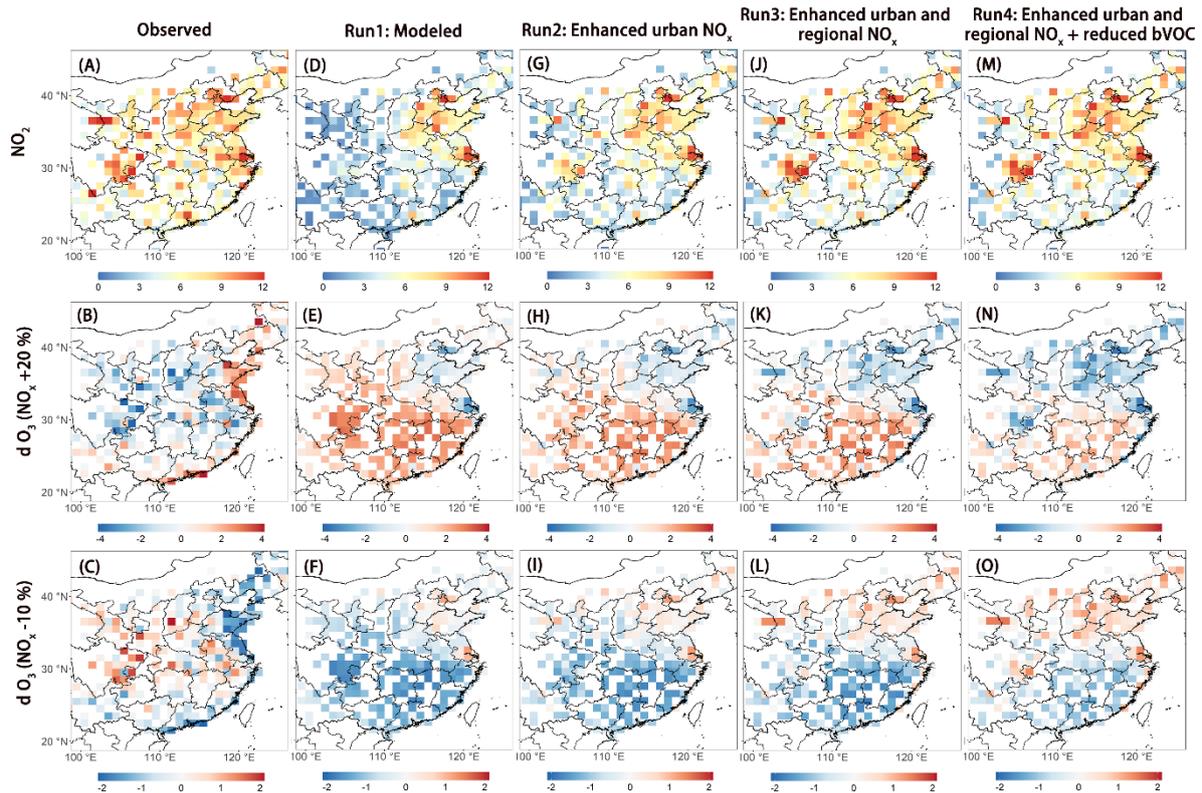

**Fig. 3**. (A) Observed surface $NO_2$ concentrations from the MEE stations in Jun-Aug 2019. The station-based measurements are averaged and re-grided to 1°×1° resolution. (B-C) Predicted responses of MDA8 $O_3$ in 2019 to $NO_2$ changes based on the lognormal fitting lines. (D) Modeled surface $NO_2$ concentrations (Run1). The modeled $NO_2$ in panel D are adjusted using the ratios of $NO_y/NO_2$ to consider the influences from reactive oxidized nitrogen compounds in the chemiluminescence analyzers. (E-F) Modeled responses of MDA8 $O_3$ in 2019 to $NO_x$ emission changes. (G-O) Similar to panels D-F, but for sensitivity simulations (Runs 2-4). The modeled $NO_2$ and $O_3$ are sampled at the locations and times of MEE surface measurements, and then averaged and re-grided to 1°×1° resolution. The 8-hour range of surface $NO_2$ measurements is selected according to the time range of MDA8 $O_3$. The unit is ppb.

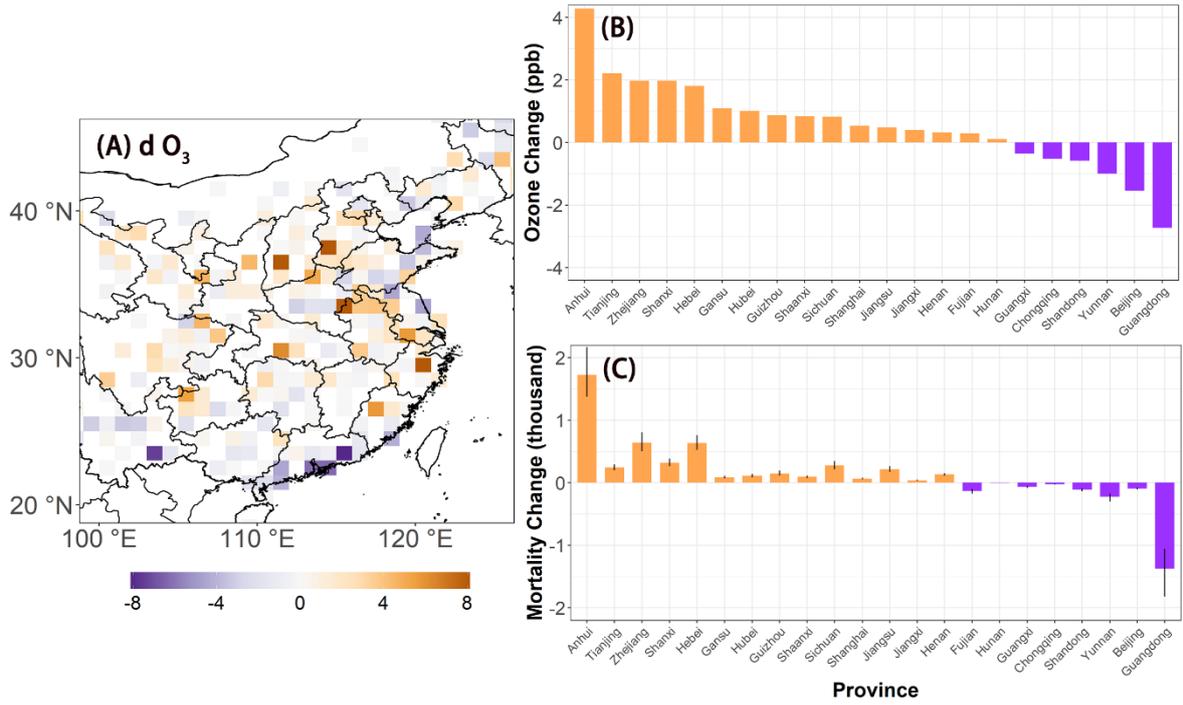

**Fig. 4.** (A-B) $NO_2$-induced urban $O_3$ changes in 2015-2020 based on the lognormal fitting method with unit ppb. (C) Changes in urban premature mortality burdens due to predicted $O_3$ changes.

# Supplemental Information

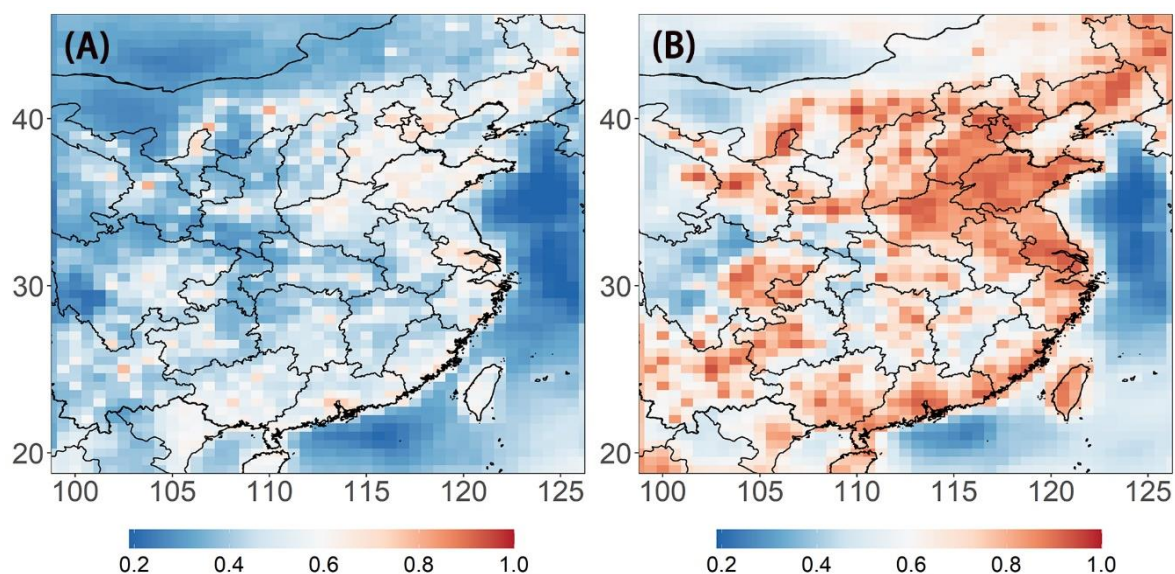

**Fig. S1.** Ratios of $NO_2/NO_y$ in GEOS-Chem simulations in Jun-Aug 2019. Here $NO_y = NO_2 + \Sigma AN + 0.95 \times PAN + 0.35 \times HNO_3$, where $\Sigma AN$ is the sum of all alkyl nitrate concentrations. (A) MDA8 based on $O_3$; (B) Daily averages.

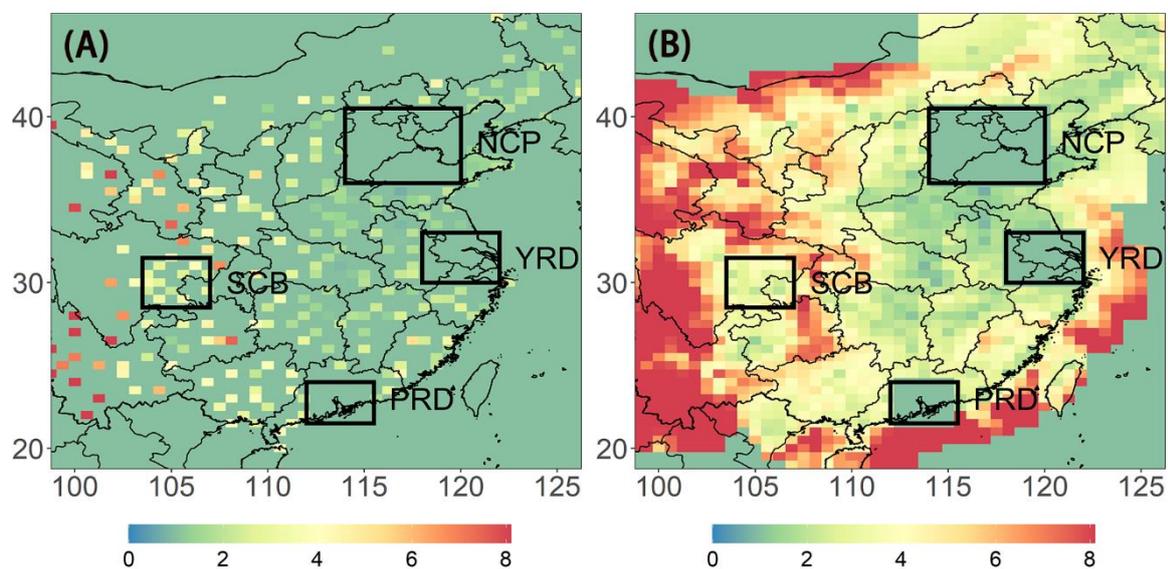

**Fig. S2.** (A) Ratios of MEE/Modeled $NO_2$ (0.5°×0.625°) over urban grids (i.e., grids have MEE urban stations). The modeled $NO_2$ are adjusted using the ratios of $NO_y/NO_2$. (B) Besides urban grids, adjusting regional background $NO_2$ based on the averaged MEE/Modeled $NO_2$ within neighboring ±1 grid domain (1.5°×1.875°); if no neighboring MEE station was found, then searching neighboring ±2 grid domain (2.5°×3.125°); the search stopped at neighboring ±4 grid domain (4.5°×5.625°).